\documentclass[prx,twocolumn,showpacs,floatfix,superscriptaddress,amsmath,amsfonts,amssymb,preprintnumbers,citeautoscript,aps]{revtex4-1}
\usepackage{graphicx, color, dcolumn, natbib, bm, amssymb, amsmath, soul}
\usepackage[table,  dvipsnames]{xcolor}
\graphicspath{{fig/}}
\usepackage[final=true]{hyperref}
\usepackage[normalem]{ulem}

\newcommand{\mn}{Mn$_{1.4}$PtSn}
\begin{document}
\title{Hybrid Bloch-N\'eel spiral states in Mn$_{1.4}$PtSn probed by resonant soft x-ray scattering}
\author{A. S. Sukhanov}\thanks{These authors contributed equally.}
\affiliation{Institut f{\"u}r Festk{\"o}rper- und Materialphysik, Technische Universit{\"a}t Dresden, D-01069 Dresden, Germany}
\email{aleksandr.sukhanov@tu-dresden.de}
\author{V. Ukleev}\thanks{These authors contributed equally.}
\affiliation{Laboratory for Neutron Scattering and Imaging (LNS), Paul Scherrer Institute (PSI), CH-5232 Villigen, Switzerland}
\affiliation{Helmholtz-Zentrum Berlin f\"ur Materialien und Energie, D-12489 Berlin, Germany}
\email{victor.ukleev@helmholtz-berlin.de}
\author{P. Vir}
\affiliation{Institut Laue-Langevin, F-38042 Grenoble, France}
\author{P. Gargiani}
\affiliation{ALBA Synchrotron Light Source, E-08290 Cerdanyola del Vall\'es, Barcelona Spain}
\author{M. Valvidares}
\affiliation{ALBA Synchrotron Light Source, E-08290 Cerdanyola del Vall\'es, Barcelona Spain}
\author{J. S. White}
\affiliation{Laboratory for Neutron Scattering and Imaging (LNS), Paul Scherrer Institute (PSI), CH-5232 Villigen, Switzerland}
\author{C. Felser}
\affiliation{Max Planck Institute for Chemical Physics of Solids, D-01187 Dresden, Germany}
\author{D. S. Inosov}
\affiliation{Institut f{\"u}r Festk{\"o}rper- und Materialphysik, Technische Universit{\"a}t Dresden, D-01069 Dresden, Germany}
\affiliation{W\"urzburg-Dresden Cluster of Excellence on Complexity and Topology in Quantum Matter\,---\,\textit{ct.qmat}, Technische Universit{\"a}t Dresden, 01069 Dresden, Germany}

\begin{abstract}

Multiple intriguing phenomena have recently been discovered in tetragonal Heusler compounds, where $D_{2d}$ symmetry sets a unique interplay between Dzyaloshinskii-Moriya (DM) and magnetic dipolar interactions. In the prototype $D_{2d}$ compound Mn$_{1.4}$PtSn, this has allowed the stabilization of exotic spin textures such as first-reported anti-skyrmions or elliptic Bloch-type skyrmions. While less attention has so far been given to the low-field spiral state, this remains extremely interesting as a simplest phase scenario on which to investigate the complex hierarchy of magnetic interactions in this materials family. Here, via resonant small-angle soft x-ray scattering experiments on high-quality single crystals of Mn$_{1.4}$PtSn at low temperatures, we evidence how the underlying $D_{2d}$ symmetry of the DMI in this material is reflected in its magnetic texture. Our studies reveal the existence of a novel and complex metastable phase, which possibly has a mixed character of both the N\'{e}el-type cycloid and the Bloch-type helix, that forms at low temperature in zero fields upon the in-plane field training. This hybrid spin-spiral structure has a remarkable tunability, allowing to tilt its orientation beyond high-symmetry crystallographic directions and control its spiral period. These results broaden the reachness of Heusler $D_{2d}$ materials exotic magnetic phase diagram and extend its tunability, thus enhancing a relevant playground for further fundamental explorations and potential applications in energy saving technologies.

\end{abstract}

\maketitle


In the past decade, noncentrosymmetric magnetic systems attracted attention of researchers due to the demonstration of topologically non-trivial spin textures stabilized by the antisymmetric Dzyaloshinskii-Moriya interaction (DMI) and their potential use in spintronics \cite{bogdanov1994thermodynamically,muhlbauer2009skyrmion,nagaosa2013topological,fert2017magnetic}. In these materials, the competition between the DMI and the symmetric exchange results in a variety of magnetic field-induced topological spin-swirling textures, such as skyrmions, anti-skyrmions, bi-skyrmions, chiral bobbers, and merons~\cite{gobel2021beyond}. The DM interaction in systems of different types is described by energy contributions linear in the first spatial derivatives of the magnetization $M$, so-called Lifshitz invariants ($L$) \cite{dzyaloshinsky1958thermodynamic,bogdanov1999stability,bogdanov1994thermodynamically}:
\begin{equation}
  L_{jk}^i = M_j \frac {\delta M_k}{\delta i} - M_k \frac {\delta M_j}{\delta i}
\end{equation}
Lifshitz invariants promote spatially modulated magnetic structures with a fixed rotation sense along specific directions. It was shown both theoretically and experimentally that the Bloch-type helices and the skyrmions arise in cubic helimagnets with the $T$ and $O$ point-group symmetries\cite{roessler2006spontaneous,muhlbauer2009skyrmion,wilhelm2011precursor}; while the N\'eel-type modulations are attributed to uniaxial systems with the $C_{nv}$ symmetries \cite{leonov2016properties,kezsmarki2015neeltype}. The $D_{2d}$ symmetry, in turn, leads to a different set of the DMI vectors.
In the crystal structure of the tetragonal $D_{2d}$ Heusler alloy \mn, the atomic layers are stacked in a way that causes the DMI vector to rotate within the $ab$-plane, such that it is parallel to the Mn-Mn bonds along the crystallographic $\langle100\rangle$ axes, and perpendicular to the bonds running along $\langle110\rangle$\cite{nayak2017magnetic,vir2019tetragonal,meshcheriakova2014large}. The symmetry also implies that the said rotation is inverted, which means that the DMI vector rotates counter-clockwise under a clockwise rotation of the atomic plane. In other words, the DMI changes its sign when the rotation by $\pi/2$ over the $c$-axis is applied. The direct consequence of this is that the Bloch-type helicoidal modulations are favored for propagation vectors along $\langle100\rangle$, whereas the propagation vectors of $\langle110\rangle$ would correspond to the N\'eel cycloids\cite{bogdanov1994thermodynamically}. Remarkably, this specific symmetry form of the DMI also results in stabilization of the antiskyrmion state in a finite applied magnetic field applied parallel to the $c$-axis \cite{nayak2017magnetic,jena2020elliptical,jena2020evolution,saha2019intrinsic,peng2020controlled}.

\begin{figure}
\includegraphics[width=0.99\linewidth]{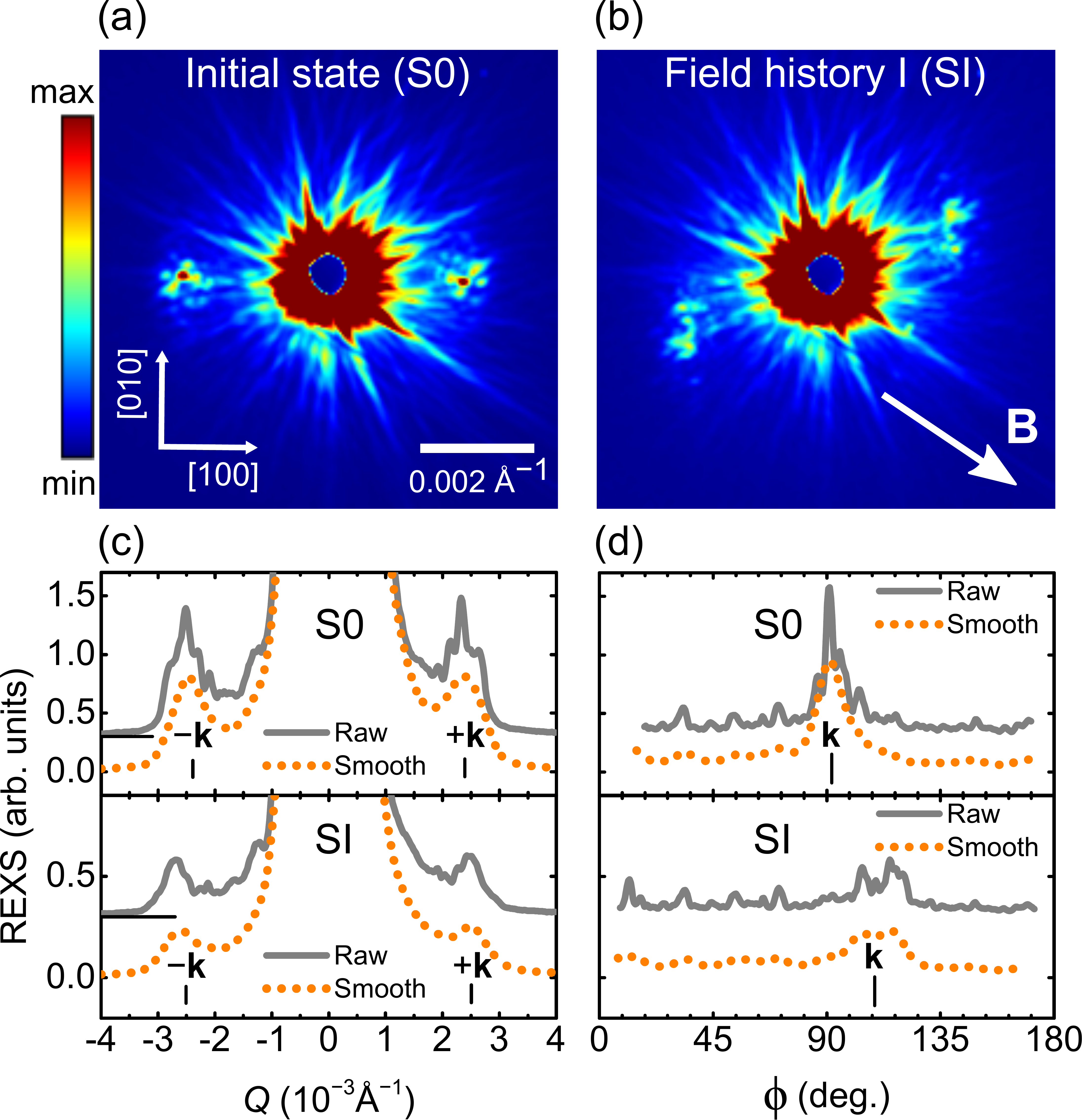}\vspace{3pt}
        \caption{REXS patterns measured at 100\,K in the (a) initial state ("S0") and (b) after the in-plane field training ("SI"). The \textit{ex-situ} training field direction is shown in the panel (b). (c) Radial and (d) azimuthal (in detector's plane) intensity profiles of the REXS patterns taken in S0 and SI states. $\phi=90^\circ$ corresponds to $\mathbf{Q||}$[$\bar{1}$00]. The solid dark-grey (dashed yellow) curves show the raw (smoothed) data. The curves are offset for clarity. The radial cuts were taken at the azimuthal angles that correspond to the maximal intensity of the Bragg peaks as rectangular stripes with a width of $\Delta Q \approx 6\cdot 10^{-4}$~\AA. The azimuthal profiles correspond to the intensity integrated over a thin ring enclosing the peaks.}
        \label{ris:fig1}
\end{figure}

The rotation sense of the spin spirals is also discerned in the internal structure of the antiskyrmion which represents a double-$k$ superposition of two proper-screw helices and two cycloids rotated by 45$^\circ$ with respect to each other. In addition to DMI, dipolar interactions play an important role for the plethora of field-induced phases in the $D_{2d}$ Heusler compound \mn~and their characteristic periodicity. The complex interplay between the anisotropic DMI and dipolar interaction leading to a variety of field-induced magnetic textures (antiskyrmions, elliptic Bloch-type skyrmions, non-topological bubbles) was first discussed in relation to Lorentz transmission electron microscopy (LTEM) studies \cite{peng2020controlled,jena2020elliptical,ma2020tunable}.

In the present study, we employ resonant small-angle x-ray scattering (REXS) to address the in-plane anisotropy of the zero-field spiral states in a thin plate of \mn. Particularly, we verify if the the spiral propagation vector can be manipulated by in-plane magnetic field history. The advantage of REXS is that it provides high resolution in momentum space and allows one to accurately determine the magnitude and the orientation of the spin-structure propagation vector of a material via Bragg diffraction\cite{okamura2017directional,okamura2017emergence,ukleev2019element,burn2019helical}. Thus, even subtle changes in the spin texture with controlled parameters (such as temperature or magnetic field) can be observed and further systematized to complement the real-space studies. By perturbing the material with an in-plane magnetic field, we demonstrate that the spin texture in \mn~can be driven in a novel metastable state, which is characterized by a tilted orientation of the spin-spiral with respect to the high-symmetry crystallographic directions. This opens up a possibility to fine-tune the spiral propagation direction within the tetragonal $ab$ plane.


A single crystal of \mn~($T_{\text{C}} = 392$~K) was grown by the flux method. The details of the crystal growth are discussed in Ref. \onlinecite{vir2019tetragonal}. The soft x-ray scattering experiment was conducted using the MARES chamber at the beamline BL29--BOREAS~\cite{barla2016design} of the ALBA synchrotron (Barcelona, Spain). Energy of the circularly polarized x-rays was tuned to Mn $L_3$ edge at $E=639$\,eV. To prepare a sample for the soft x-ray scattering measurements in transmission geometry, we cut a thin plate from a bulk single crystal using focused ion beam (FIB) milling technique. For this, we selected one of the high-quality flux-grown single crystals used in the previous studies~\cite{sukhanov2020anisotropic,vir2019tetragonal,vir2019anisotropic}. The single crystal was oriented by x-ray backscattering Laue to ensure that the crystallographic [001] axis was normal to the thin plate cut from the bulk crystal. After the FIB process, the plate with the thickness of $\sim$500~nm and the lateral dimensions of $\sim$10$\times$10~$\mu$m$^2$ was placed on a Si$_3$N$_4$ membrane and attached to it by Pt deposited on one of the corners of the plate. Prior to the sample attachment, the Si$_3$N$_4$ membrane was covered with a ca. 1\,$\mu$m-thick layer of gold on one side, and an 8~$\mu$m hole was milled through the membrane and the Au layer to allow transmission of x-rays only through the device. The sample plate was put on top of the membrane such that the center of the sample coincides with the center of the hole, which serves as the diaphragm for the incoming x-ray beam. The sample preparation process by FIB was similar to the one described in more detail in Ref. \cite{ukleev2019element}.

For all the measurements, the thin plate was oriented perpendicular to the incident x-ray beam, i.e. the crystallographic [001] axis of the sample is along the beam. The intensity of the transmitted beam was measured by a photodiode, whereas the scattering intensity was recorded by a charged-coupled device (CCD). The magnetic field was applied along the incident beam either perpendicular to the plate ($B \parallel [001]$) or in the plane of the sample ($B \perp [001]$). The latter was achieved by rotating the sample by 90~deg. around the perpendicular axis. However, this was utilized only to produce a certain field history, as the measurements cannot be conducted when the sample plate is parallel to the beam due to the geometrical constraints. Similar field-training procedures were previously utilized to study the interplay of the in-plane magnetic field and various metamagnetic textures in chiral skyrmion host FeGe \cite{ukleev2020metastable}.


Transmission REXS patterns recorded after initial zero-field cooling (ZFC) and after the in-plane field training procedures are shown in Figs.~\ref{ris:fig1}(a) and \ref{ris:fig1}(b), respectively. Evidently, the ground state helical texture [Fig.~\ref{ris:fig1}(a)] is aligned with the [100] crystallographic axis, which was previously reported as the direction along which the spin spiral propagates\cite{nayak2017magnetic,ma2020tunable,peng2020controlled}. The in-plane symmetry breaking between [100] and [010] directions in the tetragonal plane is likely originated from the FIB fabrication procedure that often results in a sample thickness gradient and tensile strain. The periodicity of the helical texture $\lambda_h=2\pi/k_s\approx 250$\,nm [Fig.~\ref{ris:fig1}(c)] corresponds well to the previous value reported for 500\,nm-thick plate from LTEM and magnetic-force microscopy (MFM) measurements \cite{ma2020tunable}. Here, we denote the initial ZFC state as ``S0".

The influence of the in-plane magnetic field training was firstly evaluated at 100\,K by rotating the sample by 90$^\circ$ over an arbitrary axis laying within its plane, such that the thin plate was parallel to the beam and the magnetic field $B$. The field of $1.1$\,T was applied and consequently driven back to zero before the sample is rotated back to its initial orientation (plate perpendicular to the beam). The result of such field training is shown Fig.~\ref{ris:fig1}(b). We denote the corresponding state as ``SI". The azimuthal orientation of the magnetic Bragg peaks is evidently affected by the procedure---the peaks were tilted from [100] towards the direction perpendicular to the training field. To quantify the observed change in the scattering patterns, radial and azimuthal slices of the REXS intensity maps for the S0 and SI states are presented in Figs.~\ref{ris:fig1}(c) and ~\ref{ris:fig1}(d), respectively.

\begin{figure}
\includegraphics[width=0.99\linewidth]{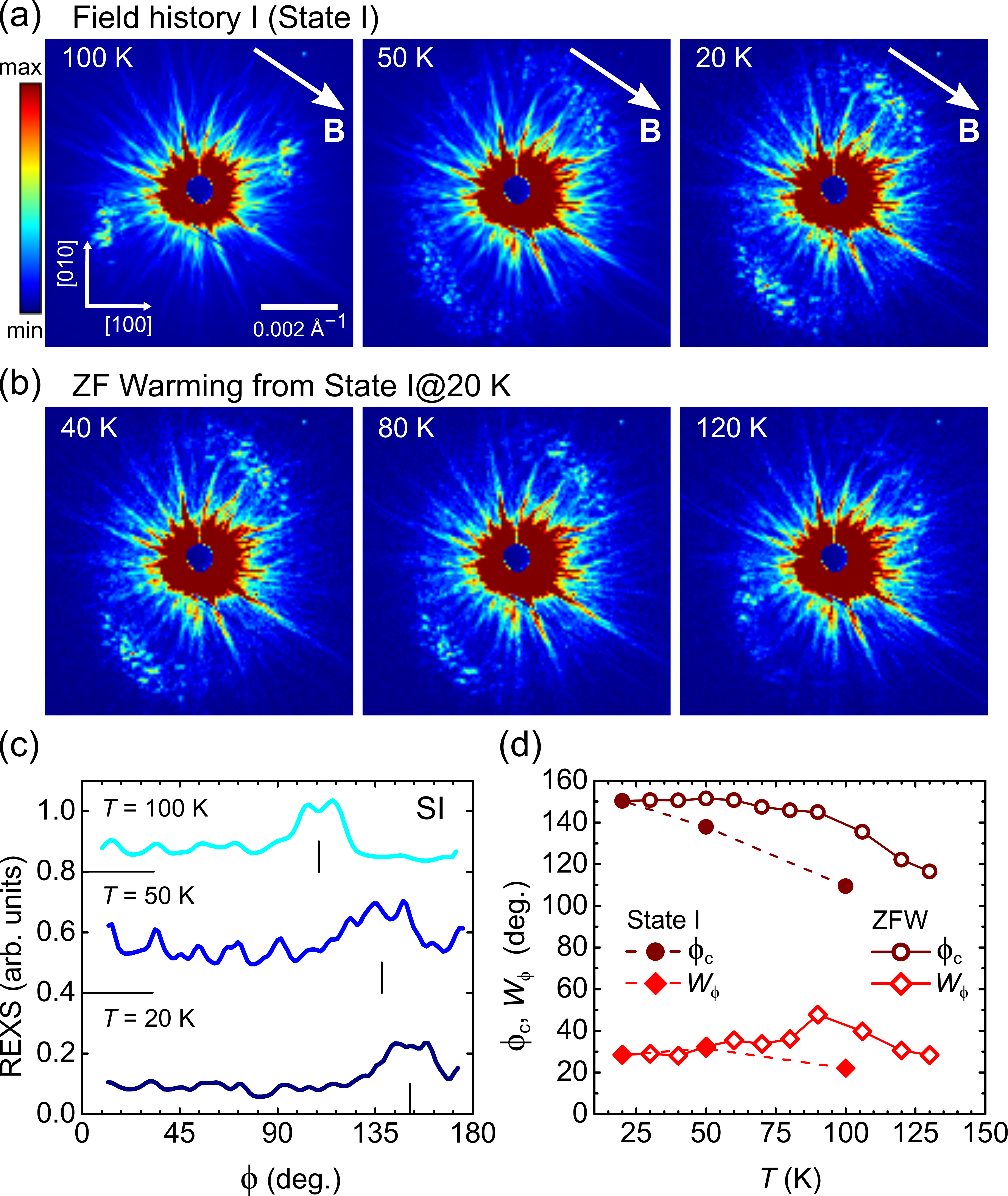}\vspace{3pt}
        \caption{(a) REXS patterns measured in the state SI at 100\,K, 50\,K and 20\,K. (b) REXS patterns measured on zero field  warming (ZFW) after preparing the SI state at 20\,K. (c) Azimuthal profiles of the REXS intensity (integrated within a thin ring enclosing the peaks) extracted from the patterns shown in (a). The data are offset for clarity. (d) Temperature dependence of the azimuthal position $\phi_{\text{c}}$ and width $W_{\phi}$ of the spiral Bragg peak in the SI state (solid symbols) and on ZFW from the SI state of 20~K (open symbols). The errorbars are smaller than the symbol size.}
        \label{ris:fig2}
\end{figure}

While the magnitude of the spin-spiral propagation vector $k_s$ is barely affected (only slightly shifted towards lower momenta $Q$) after the field training [Fig. \ref{ris:fig1}(c)], the pronounced difference occurs in the azimuthal intensity profiles that show a re-orientation of the spiral within the $ab$-plane from $\phi=90^\circ$ ($\mathbf{k}_s~||~$[100]) to $\phi \approx 110^\circ$ as a result of the in-plane field history. This direction does not correspond to any particular crystallographic axis in \mn, thus the REXS pattern reveals a metastable spin-spiral state built by the interplay of the magnetic field, the DMI and the dipolar interaction. We note that the magnetocrystalline anisotropy and the anisotropic exchange interactions may also play a role for the tilting of the spiral away from the principal crystal axes, as previously observed in chiral cubic magnet Cu$_2$OSeO$_3$ \cite{qian2018new,preissinger2021vital,leonov2020field}. Due to the symmetry of the DMI in $D_{2d}$ compounds\cite{bogdanov1994thermodynamically,meshcheriakova2014large,nayak2017magnetic}, the spiral tilted from [100] axis should be of a hybrid nature, e.g. the spin rotation plane is not the same as in the case of a proper screw helical or purely cycloidal modulations.

Further, we investigated a temperature dependence of the field training effect, as presented in Figs.~\ref{ris:fig2}(a)--\ref{ris:fig2}(d). The same field training procedure that was used to obtain the field-history state ``SI" at 100~K was repeated at 50\,K and 20\,K [Fig. \ref{ris:fig2}(a)]. It is clearly seen in Fig.~\ref{ris:fig2}(a) that the tilting angle of the spiral propagation vector after the field training increases as the temperature is lowered, such that the Bragg peaks become almost perpendicular to the direction of the training field $\mathbf{B}$ at $T=20$\,K. The in-plane rotation angle of the spiral spin texture is quantified by considering the azimuthal intensity profiles, as shown in Fig. \ref{ris:fig2}(c). As evidenced by the peak position, the applied field-training procedure at low temperatures makes the spin texture propagate close to the crystallographic [110] direction, which should constrain the spin modulations to the N\'eel-type cycloid by the symmetry, as was mentioned earlier. This possibly suggest that the N\'eel-type spiral along [110] axis is stabilized in \mn~at low temperatures by the field cycling similarly to magnetic multilayers with interfacial DMI \cite{legrand2018hybrid}. It should be noted that the propagation vector tends to rotate further beyond [110] ($\phi_{c} = 135^{\circ}$) to higher angles of $\sim$150$^{\circ}$ at the lowest temperatures. It therefore slightly deviates from [110] to orient closer to the direction that is perpendicular to the applied field (at $\phi \approx 55^{\circ}$).

Next, a zero-field warming (ZFW) procedure of the trained cycloidal state (referred to as ``SI@20~K") was carried out without introducing any further magnetic field history. The corresponding REXS patterns collected at 40\,K, 80\,K and 120\,K are shown in Fig.~\ref{ris:fig2}(b). Figure~\ref{ris:fig2}(d) summarizes the results of the temperature dependence of the spiral tilting angle and the azimuthal width of the Bragg peaks. A clear hysteretic behavior of the azimuthal center of the peak $\phi_{\text{c}}$ between the field-trained SI and the ZFW spiral states appears across the wide temperature range.

A smooth decrease of the spiral tilt angle on warming suggests the gradual restoration of the helical modulation along [100] towards higher temperatures ($T\sim150$\,K). This reflects the metastability of the low-$T$ cycloidal state that is induced by the field training but is seemingly unfavored by the DMI, for which the state with $k_{\text{s}}\parallel$[100] minimizes the energy. The spiral orientation remains pinned to its low-temperature field-induced direction on warming up to $\sim$80~K. Upon further warming, a spiral rotation back to its equilibrium becomes noticeable [Fig.~\ref{ris:fig2}(d)]. As can be seen, the pinning sites in the sample plate prevent the uniform reorientation of the spiral texture across the sample, which is reflected in the broadening of the Bragg peaks [denoted as $W_{\phi}$ in Fig.~\ref{ris:fig2}(d)], as the spin texture within the sample breaks into domains of slightly misoriented propagation vectors. The peak width becomes highest at $T \approx 90$~K, after which it decreases back evidencing more evenly distributed spiral orientations within the sample at higher temperatures.

\begin{figure}[t]
\includegraphics[width=0.99\linewidth]{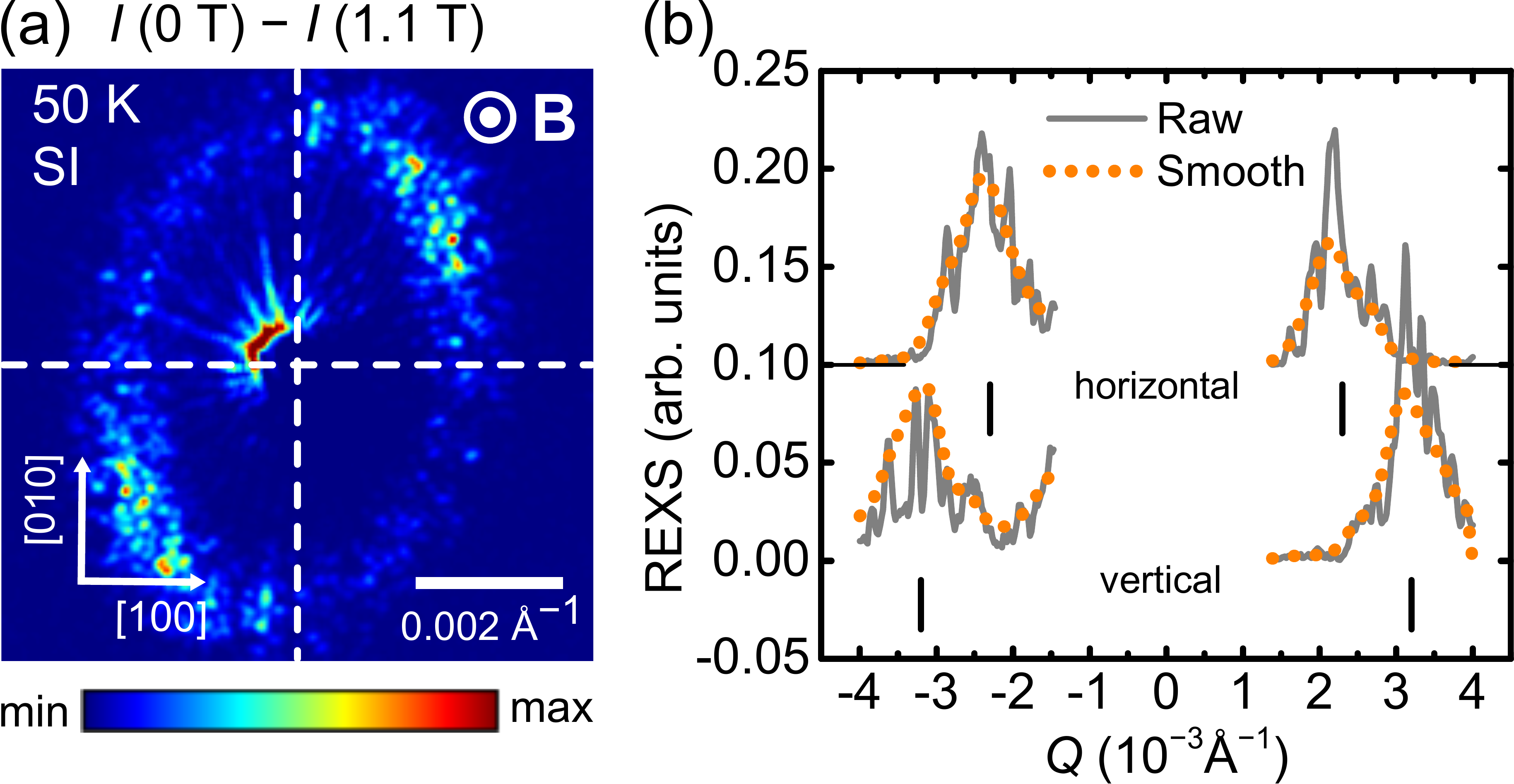}\vspace{3pt}
        \caption{(a) Zero-field REXS pattern measured at 50\,K in the SI state with the high-field background subtracted. (b) Radial profiles of the REXS intensity extracted from the horizontal (top curves) and vertical (bottom curves) cuts of the two-dimensional map. The data are offset for clarity. The smoothed signal (dotted orange line) is shown along with the raw data (solid dark-gray line).}
        \label{ris:fig3}
\end{figure}

To highlight the non-trivial in-plane REXS intensity distribution in the ``hybrid" spin-spiral state, the two-dimensional map of the REXS intensity measured at 50\,K in the SI state is shown in Fig.~\ref{ris:fig3}(a) after background subtraction. To account for the background, we collected a REXS pattern in the high (1.1\,T) out-of-plane magnetic field. Indeed, at $B=1.1$\,T applied along the $c$-axis \textit{in situ} the scattering intensity from the incommensurate satellites is suppressed as the sample undergoes the transition to the field-polarized state. Hence, the remaining signal after the background subtraction provides the enhanced view of the weak zero-field REXS intensity distribution. Figure \ref{ris:fig3}(a) demonstrates that the spiral Bragg peaks are centered at the azimuthal angles $\phi \approx 135^\circ$ and $\approx -45^\circ$ corresponding to [110] and [$\bar{1}\bar{1}$0] directions and hence the N\'eel-type cycloidal modulation. However, weak but clear REXS intensity is actually distributed across the whole $(Q_x,Q_y)$ scattering plane in a form of azimuthally-elongated Bragg tails. Moreover, the magnitude of the spiral propagation vector $k_{\text{s}}$ forms an ellipse with the long axis being parallel to [010] and the short axis being parallel to [100]. Thus, the helical spin modulations that propagate along the symmetry-equivalent axes [100] and [010] exhibit the maximal difference in the magnitude of $k_{\text{s}}$, as we show in the radial momentum-intensity cuts in Fig. \ref{ris:fig3}(b). We note that in the case of a multidomain or disordered spiral texture in a system with the $D_{2d}$ symmetry but an isotropic \textit{magnitude} of the DMI, one would expect a circular distribution of the intensity. The latter takes place when the spiral period is driven only by the competition of the DMI and the Heisenberg exchange interaction (which otherwise stabilizes the collinear state)\cite{bogdanov1994thermodynamically}. The present elliptical REXS pattern demonstrates the complex interplay between the intrinsically-anisotropic DMI in this compound and the extrinsic magnetic dipolar interaction enhanced in the thin-plate geometry of the sample. A similar mechanism was previously proposed for the formation of the elliptic skyrmions in \mn~under an out-of-plane magnetic field \cite{peng2020controlled,jena2020elliptical}. 

The detailed spin structure of the hybrid spiral states can be further examined by the direct resonant x-ray probe by exploiting circular dichroism in the reflection geometry which leaves a room for further studies using soft x-rays \cite{zhang2017direct}. Also, a REXS experiment with a vector field would be beneficial for the \textit{in-situ} observation of the spiral state under an in-plane field rotation \cite{ukleev2021signature}.


To conclude, we conducted REXS measurements of a thin-plate of \mn~at the Mn $L_3$ edge at temperatures well below the Curie temperature ($T_{\text{C}} = 392$~K). The REXS patterns collected at a low temperature after ZFC in a virgin magnetic state reveal a pair of the magnetic Bragg peaks that corresponds to the spin-spiral state with a period of $\sim$250~nm. The spiral period, as well as the orientation of the spiral propagation vector, in the ZFC state was found in agreement with the previous observation by the real-space techniques~\cite{ma2020tunable}. In order to investigate the influence of the in-plane magnetic-field history on the spin texture, we collected REXS patterns at low temperatures after a relatively high magnetic field of 1.1~T was applied to the sample within its $ab$-plane. The applied field of 1.1~T is much greater than the field needed to suppress the helical structure when the field is applied long the $c$ axis~\cite{nayak2017magnetic}. The redistribution of the zero-field REXS intensity after the in-plane field history showed that the spin-spiral structure of \mn~is not rigidly confined to the high-symmetry crystallographic axes of the material, but can instead be freely rotated in the $ab$-plane. The tilting angle is shown to depend on the sample temperature and the applied field direction. Moreover, we demonstrated that the spiral period can be controlled via the tilting angle. This is seemingly governed by the competition between the DMI and the magnetic dipolar interactions in the thin-plate geometry. This observation suggests that the in-plane orientation of the spin spiral can be utilized as another tuning parameter in the materials with the $D_{2d}$ symmetry.

\section*{Acknowledgments}

We thank Eugen Deckardt, Martin Bednarzik and Thomas Jung for their help in preparation of the membranes at PSI, and Yangkun He for his assistance in FIB nano-fabrication at MPI CPfS. Synchrotron measurements were carried out at the beamline BL-29 BOREAS at ALBA synchrotropn as a part of the proposal 2018093130. V.U. and J.S.W. acknowledge funding from the SNSF Project Sinergia CRSII5\_171003 NanoSkyrmionics. A.S.S. and D.S.I. acknowledge financial support from the German Research Foundation (DFG) under Grant No. IN 209/9-1, via the project C03 of the Collaborative Research Center SFB 1143 (project-id 247310070) at the TU Dresden and the W\"urzburg-Dresden Cluster of Excellence on Complexity and Topology in Quantum Materials — \textit{ct.qmat} (EXC 2147, Project Id 390858490). M.V. and P.G. acknowledge additional funding by grants PID2020-116181RB-C32 and FlagEra SOgraphMEM PCI2019-111908-2 (AEI/FEDER).

\bibliography{biblio}

\end{document}